\documentclass[12pt,twoside]{article}
\usepackage[utf8]{inputenc}
\usepackage{tikz}
\usepackage{varwidth} 
\usetikzlibrary{matrix, positioning}
\usepackage[hyphens]{url}

\usepackage{amsmath,graphicx,microtype,amssymb,setspace,verbatim,multirow,url,bbm,rotating,booktabs,amsthm}
\usepackage{mathrsfs}
\usepackage[font={normal}]{caption,subfig}
\usepackage{libertine}
\usepackage{color} 
\usepackage{sgame}
\usepackage{enumitem}
\usepackage[colorlinks=true,linkcolor=blue,citecolor=blue,urlcolor=blue]{hyperref}
\usepackage[nameinlink,capitalize]{cleveref}
\usepackage{bbm}

\makeatletter
	\renewcommand{\abstract}[1]{\def \@abstract {#1}}
	\newcommand{\jelcodes}[1]{\def \@jelcodes {#1}}
	\newcommand{\keywords}[1]{\def \@keywords {#1}}
	\newcommand{\thanknotes}[1]{\def \@thanknotes {#1}}
	\newcommand{\contact}[1]{\def \@contact {#1}}
	\newcommand{\shortauthor}[1]{\def \@shortauthor {#1}}
	\newcommand{\shorttitle}[1]{\def \@shorttitle {#1}}
\makeatother

\jelcodes{}
\thanknotes{}
\shortauthor{}
\shorttitle{}
\abstract{}

\usepackage[paperwidth=8.5in, paperheight=11in]{geometry}
\usepackage{setspace}
\usepackage[hang,flushmargin]{footmisc} 
\usepackage[authoryear]{natbib}
\usepackage{tikz}
\makeatletter
\usepackage{fancyhdr}
\makeatother

\newcommand\blfootnote[1]{%
  \begingroup
  \renewcommand\thefootnote{}\footnote{#1}%
  \addtocounter{footnote}{-1}%
  \endgroup
}

\usepackage{authblk}
\makeatletter
\def \maketitle { 
	\thispagestyle{empty}
	\vspace*{0.1in}
	\blfootnote{\textsc{Contact.} \@contact.  \@thanknotes\\}
	\begin{center}
	\begin{minipage}{5.2in}
	\begin{center}
	{\large {\textbf{\@title}}}
	
	\vspace{0.2in}
	
	{\textsc{\@author}}
	
	\vspace{0.2in}
	
	{\@date}
	\end{center}
	
	\ifx\@abstract\@empty
	\relax
	\else
	{\small{\textsc{Abstract.} \@abstract}}
	\fi
	
	\ifx\@keywords\@empty
	\relax
	\else
	\vspace{0.2in}
	
	{\small\textsc{Keywords.} \@keywords.}
	\fi
	
	\ifx\@jelcodes\@empty
	\relax
	\else
	{\small\textsc{JEL Codes.} \@jelcodes.}
	\fi
	
	\end{minipage}
	\end{center} }
\makeatother

\makeatletter
\def\@seccntformat#1{\csname the#1\endcsname.\ }
\makeatother

\usepackage{sectsty}
\allsectionsfont{\noindent\normalsize}
\sectionfont{\centering\normalsize\uppercase}
\paragraphfont{\textnormal}

\providecommand{\U}[1]{\protect\rule{.1in}{.1in}}

\newtheorem{theorem}{Theorem}

\newtheorem{corollary}{Corollary}

\newtheorem{lemma}{Lemma}

\newtheorem{proposition}{Proposition}

\newtheorem{assumption}{Assumption}

\usepackage{hyperref}

\usepackage{tabu}

\makeatletter
\renewcommand\@biblabel[1]{}
\makeatother

\begin{document}

\title{Selling Consumer Data under Limited Commitment}

\author{\textsc{Zihao Li*}}

\affil{*Columbia University}

\date{\today}
\abstract{A data broker repeatedly negotiates with a producer who has privately known
production costs and values consumer data for price discrimination. The broker
cannot commit to future offers. When he can offer rich menus of dataset--price
pairs, the unique equilibrium outcome immediately implements the
commitment-optimal mechanism, so limited commitment is irrelevant. When he can
offer only a single dataset--price pair at a time, by contrast, we obtain a folk
theorem. There is always an equilibrium in which the market clears immediately
at a low price, while, when the parties are sufficiently patient, any payoff
between this benchmark and the commitment payoff can be sustained. This
multiplicity rests on a property specific to data: free disposal by the producer
and zero marginal cost of provision make the efficient dataset for a given type
nonunique, leaving the broker credible flexibility over future offers.
Equilibrium multiplicity is therefore an economic prediction of the model, with
identical fundamentals supporting persistently different negotiated outcomes.
}


\shortauthor{Li}
\shorttitle{}

\contact{\textsc{zl3366@columbia.edu}}
\thanknotes{We thank Qingmin Liu for continual invaluable advice and support throughout this project. We thank Dilip Abreu, Drew Fudenberg, Faruk Gul, Alessandro Lizzeri, Alessandro Pavan, Wolfgang Pesendorfer, Evan Sadler, and Kai Hao Yang for insightful discussions. We also thank Daniel Chen, Roberto Corrao, Mira Frick, Ryota Iijima, Andrew Koh, Tianhao Liu, Yijun Liu, Daniel Luo, Teddy Mekonnen, Harry Pei, Laura Veldkamp, Yangfan Zhou, Columbia Theory Lunch, and Princeton Theory Lunch for helpful comments.}
\setcounter{page}{0}

\maketitle
    
\newpage

\section{Introduction}
Consumer data have become an important input into firms' pricing and marketing decisions. Data brokers collect information about consumers and sell it to firms, which use it to identify profitable customers and price discriminate more effectively.

Survey evidence from online data marketplaces documents substantial bilateral interaction between buyers and sellers and shows that price negotiation accounts for a large share of the effort required to complete transactions \citep{kennedy2022revisiting}. Commercial platforms explicitly accommodate such negotiations. For example, AWS Data Exchange allows providers to make customized private offers whose prices, durations, payment schedules, and contractual terms may differ from those of publicly available offers. Following a rejection, a seller can return with a new offer on different terms. Sellers in these markets therefore cannot easily commit to the terms they will offer in the future.

This paper asks how the inability to commit to future offers affects the surplus a data broker can extract. A producer who rejects an offer today anticipates that the broker may revise his terms tomorrow, while the broker cannot credibly commit to maintaining an offer that will no longer be optimal after observing a rejection. We study this question in a dynamic model of consumer-data sales. A producer has a privately known production cost, and a data broker owns information about a population of heterogeneous consumers. By purchasing a dataset, the producer gains access to the consumer profiles it contains and can price discriminate among those consumers. Lower-cost types profitably serve a larger set of consumers and therefore have a higher value for additional consumer profiles. The broker uses both dataset content and price to screen the producer's private information.

A key modeling choice is what the broker can offer at any point in the negotiation. Private negotiations are generally not observed in sufficient detail to determine whether sellers present several alternative dataset--price packages simultaneously or instead make one proposal at a time. The available evidence therefore does not single out a natural contracting protocol. We study two polar cases. In the \emph{rich-contract} environment, the broker can offer an arbitrary menu of dataset--price pairs in each period. In the \emph{simple-contract} environment, he can offer only a single dataset--price pair at a time. Comparing the two environments isolates how the richness of within-period contracting interacts with limited commitment.

As a benchmark, we characterize the static problem in which the broker can commit to a mechanism. Under our assumptions, the commitment-optimal mechanism serves every producer type, assigning different datasets and payments to different types. Lower-cost types receive larger datasets because they can profitably serve consumers with lower willingness to pay. The broker's payoff in this mechanism is the commitment benchmark against which we evaluate the dynamic game.

Our first dynamic result shows that limited commitment is irrelevant when the broker can offer sufficiently rich contracts. If arbitrary menus are available in every period, the unique equilibrium outcome implements the commitment-optimal allocation immediately, for every discount factor. The broker can screen all producer types within a single period, and because the commitment-optimal menu already clears the market, there is no gain from using delay to screen types over time. This result is not specific to data markets. It reflects a broader property of bargaining over differentiated goods: when a seller can offer a sufficiently rich menu of qualities and prices, private information can be screened within the period rather than through delay.

What is specific to data emerges when the broker is restricted to simple contracts. When only one dataset--price pair can be offered at a time, screening must occur intertemporally. One might therefore expect the standard Coasian force to erode the broker's bargaining power. Instead, we obtain equilibrium multiplicity. There is always a low-payoff equilibrium in which the broker clears the market immediately at the maximal revenue attainable from a single static offer. For any payoff between this low benchmark and the commitment benchmark, however, there is an equilibrium delivering it once the parties are sufficiently patient. High-payoff equilibria use time as a substitute for menu complexity. On the equilibrium path, the broker makes a sequence of simple offers targeted at different producer types. As the parties become patient, this sequence can approximate the screening implemented by the commitment-optimal menu. A deviation by the broker from the prescribed path triggers reversion to the low-payoff equilibrium, so the threat of losing a valuable continuation disciplines his current offers.

We interpret this construction as an endogenous form of reputation. Reputation takes the form of a self-enforcing convention about future offers. A broker who follows the prescribed bargaining path preserves the producer's expectations about continuation play, whereas a deviation destroys the associated continuation value. Reputation can therefore allow a seller who cannot formally commit to future contracts to sustain outcomes close to those attainable under commitment.

The source of this multiplicity is a distinctive economic property of data. The producer can freely discard profiles of consumers she cannot profitably serve, and because the broker already possesses the information, including additional profiles in a dataset costs nothing. Together, these features imply that the efficient dataset for a given producer type need not be unique: many distinct datasets generate the same total surplus for that type but differ in their value to other types. This nonuniqueness creates the continuation flexibility that supports equilibrium multiplicity. After the lowest-cost types have accepted earlier offers, the broker can choose among several datasets that are equally efficient for the remaining types. Because these datasets differ in their value to lower-cost types, the dataset the broker is expected to offer later affects those types' willingness to accept earlier offers. The broker can therefore vary future dataset content to alter current screening incentives without sacrificing efficiency for the remaining types.

To show that this multiplicity is not a generic consequence of quality differentiation, we study a canonical quality model in which higher quality is costly to provide and each buyer type has a unique efficient quality. In that environment, the dynamic equilibrium is essentially unique, and as players become patient, the outcome converges to efficient trade without delay and the seller's payoff is pinned down. The contrast isolates the source of multiplicity in our data market: the combination of free disposal and zero marginal cost, rather than the mere presence of a quality dimension.

The resulting equilibrium multiplicity is itself an economic prediction of the model. Consumer demand, the distribution of producer costs, and the data technology together need not pin down a unique path of negotiated prices. Outcomes may instead depend on the convention governing bargaining after rejection, and hence on seller-specific reputations, bilateral histories, or platform-specific norms. The theory thus identifies conditions under which persistent variation in negotiated outcomes can arise across sellers or platforms with identical fundamentals.

Our results relate to the Coase conjecture. In the canonical durable-goods model, the seller's inability to commit to future prices causes its bargaining power to collapse as offers become frequent. \citet{ausubel1989reputation} show that this conclusion can fail in the no-gap case, where reputational equilibria support a broad range of seller payoffs. Our model likewise generates a folk-theorem-type payoff set, but through a different mechanism. Even the highest-cost producer in our model has strictly positive willingness to pay, so we are in the gap case, in which the indifference over the timing of trade that drives their construction is absent. The relevant indifference instead concerns which efficient dataset the producer receives, and nonuniqueness of the efficient allocation plays the role that timing indifference plays in \citet{ausubel1989reputation}. Whether our simple-contract result should be viewed as a failure of the Coase conjecture nonetheless depends on what constitutes the natural per-period contract when quality is endogenous: with complete menus, the commitment outcome is implemented immediately; with one contract at a time, equilibrium selection becomes consequential. Rather than taking a stand on which protocol is the appropriate Coasian benchmark, we characterize how the predictions vary with the contracting technology.

\subsection{Relation to Literature}
The paper contributes to three literatures. First, it contributes to the literature on the sale of consumer information. The closest paper is \citet{yang2022selling}, who characterizes the optimal static mechanism through which a data broker sells market segmentations to a privately informed producer. We retain the central screening motive but introduce repeated negotiation and limited commitment. More broadly, the paper relates to work on monopoly information sales and data intermediation, including \citet{admati1986monopolistic,admati1990direct,bergemann2015selling,bergemann2018design}.

Second, the paper contributes to the literature on Coasian bargaining and durable-goods monopoly. Our analysis is related to the canonical results of \citet{FLT1985} and \citet{gul1986foundations}, and especially to the reputational equilibria of \citet{ausubel1989reputation}. As discussed above, our contribution relative to \citet{ausubel1989reputation} is to identify a different primitive source of continuation flexibility, one that operates in the gap case. For differentiated products, \citet{wang1998bargaining} shows that allowing an uninformed party to offer menus of quality contracts can generate immediate screening, closely paralleling our rich-contract benchmark. Related work includes \citet{hahn2006damaged,nava2019differentiated}. Our comparison between rich and simple contracts isolates the distinction between screening through simultaneous contractual alternatives and screening through a sequence of offers over time.

Third, our work contributes to the literature on nonlinear pricing \citep[e.g.,][]{mussa1978monopoly, maskin1984monopoly, wilson1993nonlinear} and mechanism design with ex-post moral hazard \citep[e.g.,][]{laffont1986using, carbajal2013mechanism, strausz2017theory, gershkov2021theory}. Among more recent contributions, \citet{corrao2023nonlinear} study nonlinear pricing for goods whose use generates revenue but is subject to free disposal. Relatedly, \citet{dall2026screening} study nonlinear screening for digital goods when replication and quality degradation are costless, showing how this nonseparable production technology changes the standard quality-distortion logic of \citet{mussa1978monopoly}; our focus is instead on how a similar cost structure interacts with limited commitment in dynamic bargaining. To our knowledge, we are the first to study dynamic screening with ex-post moral hazard under limited commitment.
\section{Model}
\label{sec:model}
\subsection{Primitives}
\paragraph{\textbf{Consumers.}}
There is a unit mass of consumers, each with unit demand and heterogeneous
valuations. Consumer valuations, denoted by $v$, are distributed according to
the cumulative distribution function $F(\cdot)$. Consumers are passive price
takers.

We assume that $F(\cdot)$ is differentiable and admits a density $f(\cdot)$ on
the compact interval $[\underline v,\overline v]$ that is bounded above and
away from zero: for some constants $m$ and $M$,
\[
0<m\leq f(v)\leq M
\qquad
\text{for all }v\in[\underline v,\overline v].
\]

\paragraph{\textbf{Producer.}}
The producer (she) sells a single product to these consumers but initially has
no direct means of contacting them. She therefore seeks to purchase consumer
data---for example, phone numbers or email addresses---from a data broker (he).
The producer is characterized by a privately known production cost
\[
c\in[\underline c,\overline c],
\qquad
0\leq \underline c<\overline c<\overline v.
\]
Production costs are distributed according to the cumulative distribution
function $G(\cdot)$.

We assume that $G(\cdot)$ is differentiable and admits a density $g(\cdot)$
satisfying, for the same constants $m$ and $M$,
\[
0<m\leq g(c)\leq M
\qquad
\text{for all }c\in[\underline c,\overline c].
\]
We further assume that the virtual cost
\[
\phi(c):=c+\frac{G(c)}{g(c)}
\]
is strictly increasing in $c$.

\paragraph{\textbf{Dataset.}}
Let $L^1(F)$ denote the space of $F$-integrable functions on
$[\underline v,\overline v]$, modulo equality $F$-almost everywhere
($F$-a.e.). A \emph{dataset} is an element of
\[
\mathcal B
:=
\left\{
B\in L^1(F):
B(v)\in\{0,1\}\quad F\text{-a.e.},
\qquad
B(v)=1\quad F\text{-a.e. on }\{v\geq\overline c\}
\right\}.
\]
Thus $B(v)=1$ means that consumer profile $v$ is included in the dataset, with
all such statements understood $F$-a.e. The second requirement imposes a lower
bound on dataset size: every feasible dataset contains the profiles in
\[
[\overline c,\overline v],
\]
which every producer type can profitably serve. This restriction ensures that
every dataset has strictly positive value to every type, which we use below to
order types.

We endow $\mathcal B$ with the metric
\[
d_{\mathcal B}(B,B')
:=
\|B-B'\|_{L^1(F)}
=
\int_{\underline v}^{\overline v}
|B(v)-B'(v)|\,\mathrm dF(v).
\]
The set $\mathcal B$ is a closed subset of the separable Banach space
$L^1(F)$ and is therefore Polish.

For a Borel set $D\subseteq[\underline v,\overline v]$, we identify $D$ with
the $L^1(F)$ equivalence class of its indicator $\mathbbm 1_D$. In particular,
expressions such as
\[
[\overline c,\overline v]\subseteq B
\qquad\text{and}\qquad
B=[x,\overline v]
\]
are understood in this $F$-a.e. sense. For $B,B'\in\mathcal B$, we write
$B\supseteq B'$ whenever
\[
B(v)\geq B'(v)
\qquad
F\text{-a.e.}
\]
If $B\supseteq B'$, then $B\setminus B'$ denotes the element of $L^1(F)$
represented by $B-B'$. We also use the shorthand
\[
\int_B\psi(v)\,\mathrm dF(v)
:=
\int_{\underline v}^{\overline v}
B(v)\psi(v)\,\mathrm dF(v).
\]

Given a dataset $B$, the producer can perfectly price discriminate among the
consumers whose profiles the dataset contains and obtains gross utility
\[
u(B,c)
:=
\int_{\underline v}^{\overline v}
B(v)(v-c)^+\,\mathrm dF(v)
=
\int_B(v-c)^+\,\mathrm dF(v).
\]
The map $(B,c)\mapsto u(B,c)$ is jointly Lipschitz continuous:
\[
|u(B,c)-u(B',c')|
\leq
(\overline v-\underline c)d_{\mathcal B}(B,B')
+
|c-c'|.
\]
Moreover, because every feasible dataset contains
$[\overline c,\overline v]$ and $F(\overline c)<1$ (since
$\overline c<\overline v$), the function $u(B,\cdot)$ is strictly decreasing
for every $B\in\mathcal B$. We therefore order producer types by willingness
to pay: $\overline c$ is the \emph{lowest} type, and lower-cost producers are
\emph{higher} types.

For a Polish space $S$, let $\Delta(S)$ denote the set of Borel probability
measures on $S$, equipped with the topology of weak convergence. Lotteries
over datasets appear in the commitment benchmark, and it is convenient to
summarize a lottery by its mean dataset. For any dataset lottery
$\mu\in\Delta(\mathcal B)$, define its barycenter by the Bochner integral
\[
\overline B_\mu
:=
\int_{\mathcal B}B\,\mathrm d\mu(B)
\in L^1(F).
\]
Since $\mathcal B$ is bounded in $L^1(F)$, this integral is well-defined. It
satisfies
\[
0\leq\overline B_\mu(v)\leq1
\]
$F$-a.e. and
\[
\overline B_\mu(v)=1
\]
$F$-a.e. on $\{v\geq\overline c\}$. Fubini's theorem gives
\[
\int_{\mathcal B}u(B,c)\,\mathrm d\mu(B)
=
\int_{\underline v}^{\overline v}
\overline B_\mu(v)(v-c)^+\,\mathrm dF(v).
\]

\begin{assumption}[No Exclusion under a Single Dataset]
\label{ass:no-exclusion}
For every $c\in[\underline c,\overline c]$,
\[
\int_{v\geq c}
\left(
v-c-\frac{G(c)}{g(c)}
\right)
\,\mathrm dF(v)
\geq0.
\]
\end{assumption}

\cref{ass:no-exclusion} is a no-exclusion condition for the single-dataset
problem. For every producer type, it requires the virtual surplus generated by
providing all consumer profiles that the producer can profitably serve to be
nonnegative.

Only the condition at $c=\overline c$,
\[
\int_{v\geq\overline c}
\left(
v-\overline c-\frac{1}{g(\overline c)}
\right)
\,\mathrm dF(v)
\geq0,
\]
is needed for the unrestricted static screening benchmark in
Section~\ref{sec:commitment}. The full strength of \cref{ass:no-exclusion} is
used in the analysis of the simple-contract environment to rule out
profitable exclusion deviations under the continuation strategies that we
construct.\footnote{The condition at $\overline c$ alone does not imply the
condition at interior types: lowering $c$ raises the integrand pointwise but
also adds the profiles $v\in[c,\phi(c))$, whose virtual surplus is negative.}

Section~\ref{sec:commitment} studies two static benchmarks: an unrestricted
mechanism and a mechanism restricted to a single dataset offered to all types.
Under \cref{ass:no-exclusion}, neither benchmark excludes any type: the
optimal unrestricted mechanism serves every producer type, and the optimal
single-dataset mechanism also clears the market. The unrestricted mechanism
nevertheless assigns different datasets to different types and uses dataset
content as a screening instrument.

\subsection{Timing and Solution Concept}

Let $P\subset\mathbb R$ be a compact interval of feasible payments, large
enough to contain every payment that arises in the analysis. The contract
space is
\[
\mathcal X:=\mathcal B\times P,
\]
equipped with the metric
\[
d_{\mathcal X}\big((B,p),(B',p')\big)
:=
d_{\mathcal B}(B,B')
+
\min\{1,|p-p'|\}.
\]
Because $\mathcal B$ is Polish, so is $\mathcal X$.

Let $\mathcal K(\mathcal X)$ denote the set of all nonempty compact subsets of
$\mathcal X$, equipped with the Hausdorff metric induced by
$d_{\mathcal X}$. The space $\mathcal K(\mathcal X)$ is Polish. The two offer
spaces are
\[
O^{\mathrm{pair}}
:=
\bigl\{\{x\}:x\in\mathcal X\bigr\},
\qquad
O^{\mathrm{menu}}
:=
\mathcal K(\mathcal X).
\]
In the rich-contract environment, an offer is a nonempty compact menu of
contracts; in the simple-contract environment, an offer is a single
dataset--price pair. Throughout, $O$ denotes either $O^{\mathrm{pair}}$ or
$O^{\mathrm{menu}}$, depending on the environment under consideration. When
$O=O^{\mathrm{pair}}$, we identify the singleton $\{(B,p)\}$ with the pair
$(B,p)$ itself.

Time is discrete and indexed by $t=0,1,\ldots$, with common discount factor
$\delta\in[0,1)$. In each period $t$ in which trade has not yet occurred, the
broker makes an offer $o_t\in O$. Upon observing $o_t$, the producer chooses
an action from
\[
o_t\cup\{\emptyset\},
\]
where choosing $(B_t,p_t)\in o_t$ constitutes acceptance and $\emptyset$
denotes rejection. Compactness of $o_t$ and continuity of
$(B,p)\mapsto u(B,c)-p$ ensure that a best contract exists in every menu.

Acceptance terminates the game. If the producer accepts $(B_t,p_t)\in o_t$ in
period $t$, her payoff is
\[
\delta^t\bigl[u(B_t,c)-p_t\bigr],
\]
and the broker's payoff is
\[
\delta^t p_t.
\]
If the producer never accepts, both players obtain zero.

Because only rejection leads to another period, a nonterminal history at the
beginning of period $t$ can be identified with
\[
h_t=(o_0,\ldots,o_{t-1})\in H_t:=O^t,
\qquad
H_0:=\{\varnothing\}.
\]
Let
\[
H:=\bigsqcup_{t=0}^{\infty}H_t.
\]
The producer's history after observing the current offer is
\[
\widehat h_t:=(h_t,o_t)\in\widehat H_t:=H_t\times O,
\qquad
\widehat H:=\bigsqcup_{t=0}^{\infty}\widehat H_t.
\]
All of these spaces carry the Borel $\sigma$-algebras of their product and
disjoint-union topologies.

A broker behavioral strategy is a sequence of Borel stochastic kernels
\[
\sigma_t(\cdot\mid h_t)\in\Delta(O),
\qquad
h_t\in H_t,
\qquad
t\geq0.
\]
A producer behavioral strategy is a sequence of Borel stochastic kernels
\[
\alpha_t(\cdot\mid h_t,o_t,c)
\in
\Delta\bigl(\mathcal X\cup\{\emptyset\}\bigr),
\]
such that
\[
\operatorname{supp}
\alpha_t(\cdot\mid h_t,o_t,c)
\subseteq
o_t\cup\{\emptyset\}
\]
for every $(h_t,o_t,c)$. A behavioral strategy is \emph{pure} when each of its
kernels is degenerate. We suppress the time index and write $\sigma(h_t)$ and
$\alpha(\widehat h_t,c)$ when it is clear from context.

A belief system is a sequence of Borel maps
\[
\beta_t:H_t\to
\Delta\bigl([\underline c,\overline c]\bigr),
\]
where $\beta_t(h_t)$ is the broker's belief about the producer's type
conditional on reaching $h_t$ without trade. It is useful to define the set of
active types and its highest-cost element, that is, the lowest active type:
\[
A(h_t):=\operatorname{supp}\beta_t(h_t),
\qquad
\overline c(h_t):=\sup A(h_t).
\]

A \emph{Perfect Bayesian Equilibrium} is an assessment
$(\sigma,\alpha,\beta)$ satisfying:
\begin{enumerate}
\item[(i)] \emph{Sequential rationality:} at every history, the broker's and
producer's continuation strategies maximize their respective expected
discounted payoffs given the other player's continuation strategy and the
beliefs;
\item[(ii)] \emph{Consistency:} $\beta$ is obtained from the prior $G$ and the
strategy profile by Bayes' rule whenever possible.
\end{enumerate}
We suppress the belief system and write $(\sigma,\alpha)$ for an equilibrium
whenever the associated beliefs are clear.

\section{Benchmarks}
\label{sec:commitment}

\subsection{Static Optimal Mechanism}

We first study the commitment benchmark, in which the data broker commits to a
direct mechanism that assigns each producer type a probability of trade, a
lottery over datasets, and a transfer. Formally, a mechanism $\mathcal M$ is a
mapping
\[
\mathcal M:
[\underline c,\overline c]
\to
[0,1]\times\Delta(\mathcal B)\times\mathbb R.
\]
For a type $c$, denote the components by $(q(c),B(c),p(c))$, where $q(c)$ is
the probability of trade, $B(c)\in\Delta(\mathcal B)$ is the dataset lottery
conditional on trade, and $p(c)$ is the expected transfer.

A mechanism satisfies \emph{incentive compatibility} if
\[
q(c)\,
\mathbb E\bigl[u(B(c),c)\bigr]
-
p(c)
\geq
q(c')\,
\mathbb E\bigl[u(B(c'),c)\bigr]
-
p(c'),
\qquad
\forall c,c'\in[\underline c,\overline c].
\]
It satisfies \emph{individual rationality} if
\[
q(c)\,
\mathbb E\bigl[u(B(c),c)\bigr]
-
p(c)
\geq0,
\qquad
\forall c\in[\underline c,\overline c].
\]
The broker's problem is
\[
\max_{\mathcal M}
\int_{\underline c}^{\overline c}
p(c)g(c)\,\mathrm dc
\]
subject to incentive compatibility and individual rationality.

\begin{proposition}[Commitment-Optimal Mechanism]
\label{prop:yang}
The commitment-optimal mechanism trades with every type with probability one
and assigns each producer type $c$ the deterministic dataset
\[
B^*(c)
=
\left[
\min\left\{
c+\frac{G(c)}{g(c)},
\overline c
\right\},
\overline v
\right].
\]
Each type $c$ pays
\[
p^*(c)
=
u\bigl(B^*(c),c\bigr)
-
\int_c^{\overline c}
\int_{B^*(s)}
\mathbbm 1_{\{v\geq s\}}
f(v)\,\mathrm dv\,\mathrm ds.
\]
We denote the broker's payoff in this mechanism by $\pi(G)$.
\end{proposition}

The commitment-optimal mechanism uses dataset content to screen the producer's
private cost. Lower-cost types receive weakly larger datasets because they can
profitably serve consumers with lower willingness to pay. The cutoff
\[
c+\frac{G(c)}{g(c)}
\]
has the usual virtual-cost interpretation: consumer profiles with values below
the producer's virtual cost are excluded to reduce information rents. Every
producer type is nevertheless served, so the distortion operates through
dataset content rather than through the decision to trade.

The commitment payoff $\pi(G)$ is also an upper bound on the broker's revenue
in the dynamic game. Any dynamic outcome induced by a broker strategy and a
producer best response can be reduced to a static incentive-compatible and
individually rational mechanism by treating discounted trade probabilities as
allocation probabilities, the distribution of the dataset received conditional
on trade as the dataset lottery, and discounted payments as transfers
\citep[e.g.,][]{ausubel1989direct}.

\begin{corollary}
\label{cor:yang}
No broker strategy, coupled with a producer best response, can yield the broker
a payoff strictly exceeding $\pi(G)$.
\end{corollary}

\subsection{The Rich-Contract Environment}
\label{sec:rm}

We now turn to the dynamic game with rich contracts.

\begin{proposition}
\label{prop:menu}
Fix any $\delta<1$. If $O=O^{\mathrm{menu}}$, then in every equilibrium the
commitment-optimal allocation and transfers are implemented at $t=0$: the
market clears immediately and the broker obtains the commitment payoff
$\pi(G)$.
\end{proposition}
The proof is analogous to that of \cite{Jeffery} and is therefore omitted.

\cref{prop:menu} shows that limited commitment is irrelevant when the broker can
screen producer types within a period. The commitment-optimal mechanism already
serves every type, so by offering the corresponding menu at $t=0$ the broker
screens all types within a single period. Delay therefore has no screening
value.

The proposition isolates the source of the dynamic problem studied below. A
loss of bargaining power does not arise from limited commitment alone, but from
its interaction with restrictions on the contracts available within a period.
As noted in the introduction, this conclusion is not specific to data
\citep{wang1998bargaining}; the distinctive role of data emerges in the
simple-contract environment, to which we now turn.

\section{Main Results}
\label{sec:coase}

We now turn to the simple-contract environment,
$O=O^{\mathrm{pair}}$, in which the broker can offer only one
dataset--price pair in each period. Define
\[
u_{\min}
:=
\int_{\overline c}^{\overline v}
(v-\overline c)f(v)\,\mathrm dv
\]
as the gross utility of the lowest producer type $\overline c$, which is
independent of the dataset: because every feasible dataset contains
$[\overline c,\overline v]$,
\[
u(B,\overline c)=u_{\min}
\]
for every $B\in\mathcal B$. Let $SE(\delta)$ denote the set of broker
equilibrium payoffs when the discount factor is $\delta$.

\begin{theorem}[A Folk Theorem]
\label{thm:folk}
Suppose $O=O^{\mathrm{pair}}$. For every $\epsilon>0$, there exists
$\underline\delta<1$ such that, for every
$\delta\geq\underline\delta$,
\[
[u_{\min},\,\pi(G)-\epsilon]
\subseteq
SE(\delta).
\]
\end{theorem}

\Cref{thm:folk} shows that restricting the broker to one contract at a time
does not simply lower his payoff from the commitment benchmark to a new,
uniquely determined level. Instead, the restriction generates substantial
equilibrium multiplicity. When players are sufficiently patient, the same
economic primitives support broker payoffs from $u_{\min}$ up to arbitrarily
close to the commitment payoff $\pi(G)$.

The theorem has two economic implications. First, simple contracts can make
limited commitment consequential: some equilibria give the broker substantially
less than he obtains under commitment. Second, simple contracts do not prevent
the broker from approximately attaining the commitment payoff. A sequence of
simple offers can substitute for a simultaneous screening menu, but whether it
does so depends on continuation play.

The proof builds on the following low-payoff equilibrium.

\begin{proposition}
\label{thm:Coase}
For every discount factor $0\leq\delta<1$, there exists an equilibrium in which
the broker clears the market at $t=0$ by offering
\[
\bigl([\underline v,\overline v],u_{\min}\bigr)
\]
and obtains revenue $u_{\min}$.
\end{proposition}

\cref{thm:Coase} identifies the low-payoff equilibrium that serves as the
punishment supporting \cref{thm:folk}. When the broker can offer only one
dataset--price pair at a time, he cannot, within a single period, serve the
lowest type and charge higher types different prices. The proposition shows
that the resulting low market-clearing payoff can be sustained for every
$\delta<1$, rather than only in the patient limit.

The key issue is whether, after a deviation by the broker, he can profitably
use subsequent changes in dataset content to screen producer types. The
equilibrium construction rules out this possibility by requiring that the
dataset component of the deviating offer be preserved along the relevant
continuation path. Conditional on a fixed dataset, the continuation problem
reduces to pricing that dataset to the remaining producer types.
\cref{ass:no-exclusion} ensures that the relevant deviations cannot profitably
exclude types.

Why can such persistence be sequentially rational? The answer lies in the
efficiency structure of datasets. Supplying additional consumer profiles is
costless to the broker, and the producer can simply discard profiles she
cannot profitably serve. As a result, any dataset that contains all profiles
the lowest remaining type can profitably serve is efficient for that type, so
the efficient dataset for that type is not unique.

This contrasts with the canonical quality model in which higher quality is
costly to provide. Once high
types have traded, the seller typically strictly prefers to reduce quality
toward the level efficient for the remaining low type. Anticipating this
deterioration, high types may be willing to pay for higher quality earlier,
giving the seller an intertemporal screening instrument. In our data
environment, by contrast, free disposal by the producer and zero marginal
delivery cost for the broker make efficient provision nonunique, so efficiency
alone does not pin down the broker's continuation offer. The broker can
therefore credibly preserve dataset content after higher types have traded.

\cref{thm:Coase} relies on \cref{ass:no-exclusion}. Without this condition,
exclusion may be profitable in the single-dataset continuation problems used
in the construction, so the strategy profile described above need not be
sequentially rational. We do not characterize the equilibrium set in that
case.\footnote{The usual skimming property need not hold in our environment.
Consequently, continuation play cannot in general be summarized by a single
cutoff type, and standard recursive arguments from durable-goods monopoly do
not apply directly. The proof of \cref{thm:Coase} instead constructs an
equilibrium explicitly and verifies sequential rationality after every
history.}

We now return to \cref{thm:folk}, whose proof uses the low-payoff equilibrium
of \cref{thm:Coase} as a credible punishment. The high-payoff construction
uses time as a substitute for menu complexity. On the equilibrium path, the
broker selects a sequence of cutoff types $\{c_t\}$ and, in period $t$, offers
the dataset
\[
\left[
\min\left\{
c_t+\frac{G(c_t)}{g(c_t)},
\overline c
\right\},
\overline v
\right],
\]
which is the dataset assigned to type $c_t$ by the commitment-optimal
mechanism. The price $p_t$ makes type $c_t$ indifferent between accepting
immediately and waiting for the next offer, and all higher types accept. As
$\delta\to1$ and the distance between successive cutoff types becomes small,
this sequence of simple offers approximates the allocations and transfers of
the commitment-optimal menu.

After any deviation by the broker, continuation play switches to the
low-payoff equilibrium of \cref{thm:Coase}. The prospect of losing
continuation revenue makes it optimal for the broker to adhere to the
prescribed screening path. The construction therefore admits an
endogenous-reputation interpretation. The broker's ``reputation'' consists of
the self-enforcing expectation that he will continue along a particular path
of offers. Adherence preserves the favorable continuation, whereas a deviation
triggers a less favorable one.

As emphasized in the introduction, the multiplicity in \cref{thm:folk} should
therefore be viewed as an economic prediction rather than merely an
equilibrium-selection problem. The primitives determine a set of sustainable
outcomes but need not pin down a unique outcome within that set. Bargaining
conventions, bilateral histories, or seller-specific reputations can select
among these equilibria, generating different prices and broker profits even
when preferences and technologies are held fixed, and the multiplicity
disappears under the richer contracting technology of Section~\ref{sec:rm}.

\section{Discussion}
\label{sec:discussion}

\subsection{Contracting Protocols and Empirical Interpretation}

The rich-contract and simple-contract environments should be interpreted as
two polar cases of the broker's within-period contracting capacity. In the
rich-contract environment, the producer can choose among several
dataset--price pairs at the same history. In the simple-contract environment,
only one such pair is available, so any additional screening must occur
through subsequent offers.

As noted in the introduction, existing evidence does not identify a canonical
protocol for commercial data negotiations. AWS Data Exchange allows providers to extend private offers whose
prices, durations, payment schedules, and other contractual terms may differ
from publicly available terms.\footnote{See the AWS Data Exchange
and AWS Marketplace documentation on private offers,
\cite{awsdataexchange}.} The complete sequence of alternatives
exchanged during such private negotiations is generally not publicly observed.
The available evidence therefore does not select between our rich-contract and
simple-contract environments.

This opacity also motivates limited commitment. An accepted contract is
binding, but before acceptance a seller engaged in bilateral negotiation can
revise a proposal after it is rejected. When deciding whether to accept the
current offer, the producer therefore has reason to anticipate how the broker
will alter dataset content and prices in the future. Additional evidence is consistent with the prevalence of bilateral pricing.
\citet{azcoitia2023price} collect information on more than 210,000 commercial
B2B data products from more than 2,000 sellers. Only about 4,200 products in
their sample disclose a price; many other products are either free or offered
without a fixed price and instead require direct contact with the seller.

We do not interpret these findings as direct evidence for the equilibrium
mechanism in \cref{thm:folk}. Seller identity may matter because of data
quality, reliability, legal compliance, or repeated relationships, and observed
price dispersion may reflect unobserved product heterogeneity. The evidence
instead establishes that bilateral and nontransparent contracting is
institutionally prevalent. Our model provides one theoretical channel through
which histories and bargaining conventions can matter in such an environment. A direct empirical test would require information about negotiation histories:
which alternatives are available at each stage, how prices and dataset content
change following rejection, and whether departures from customary bargaining
paths systematically change subsequent terms. Existing marketplace data
generally do not contain this information.

A final remark concerns the assumption that acceptance terminates the
relationship, which is also important for our results. If data access were
instead provided through renewable subscriptions and contractual terms could
be renegotiated after adoption, the rich-contract result (\cref{prop:menu}) would then no
longer apply directly; see, for example, \citet{strulovici2017contract}.

\subsection{Equilibrium Multiplicity as a Market Feature}

A natural concern is that the equilibrium multiplicity in \cref{thm:folk}
weakens the predictive content of the model. We take the opposite view. The
equilibrium correspondence is itself an economic object: the model predicts
that, in the simple-contract environment, preferences and technology need not
uniquely determine negotiated prices and broker profits.

In particular, markets with sufficiently patient participants, the same
consumer demand, the same distribution of producer costs, and the same data
technology can support substantially different outcomes depending on
continuation behavior. A bargaining convention governing what follows
rejection, a seller-specific reputation for adhering to a particular sequence
of offers, or a platform-specific norm can therefore become an economically
relevant state variable even though it is not part of the underlying
technology.

The model consequently generates comparative predictions about when
equilibrium selection should matter. Markets in which sellers can offer rich
within-period menus should behave like the commitment benchmark. Markets in
which negotiation proceeds through one proposal at a time can display greater
dependence on bargaining histories and conventions. Moreover, as the next
subsection shows, even simple contracts do not generate comparable
multiplicity when efficient provision is unique.

\subsection{Why Datasets Are Different}
\cref{thm:folk} relies on a structural property of datasets. Consider a
producer with cost $c$. She can profitably serve every consumer with
$v\geq c$ and can discard every profile with $v<c$. Hence any dataset
satisfying
\[
[c,\overline v]\subseteq B
\]
is efficient for that type. Additional profiles below $c$ do not reduce her
payoff, and supplying them costs the broker nothing. Efficient use of a
dataset is therefore unique, but efficient \emph{provision} is not.

This nonuniqueness is strategically important. Two datasets can be equally
efficient for the lowest remaining type but differ in their value to higher
types. The broker can therefore vary the dataset offered in the continuation
in ways that affect the incentives of other types without sacrificing surplus
for the lowest remaining type. This is the flexibility used by the
constructions in Section~\ref{sec:coase}.

The same force is absent in the canonical quality model in which higher
quality is costly to provide. The following benchmark makes the contrast
explicit.

\begin{proposition}[Unique-Quality Benchmark]
\label{prop:unique}
Consider the same simple-contract timing with two buyer types,
\[
\theta_1>\theta_2>0,
\]
with prior probabilities $q_i>0$ satisfying $q_1+q_2=1$. A buyer of type
$\theta$ obtains gross utility
\[
u(x,\theta)=\theta x
\]
from quality $x\in[\underline x,\overline x]$, while the seller's payoff from
a quality--price pair $(x,p)$ is
\[
p-\frac{x^2}{2}.
\]
Assume
\[
[\theta_2,\theta_1]
\subseteq
[\underline x,\overline x].
\]
For every $\delta\in[0,1)$, the equilibrium outcome is essentially unique and
the seller's equilibrium payoff is unique. Moreover, as $\delta\to1$, delay
vanishes and each type receives her efficient quality,
\[
x^e(\theta_i)=\theta_i,
\qquad
i=1,2.
\]
\end{proposition}
Total surplus in this benchmark is
\[
\theta x-\frac{x^2}{2},
\]
which is strictly concave in $x$. Each buyer type therefore has a unique
efficient quality. Once only the low type remains, sequential rationality pins
down the seller's continuation quality: the seller strictly prefers the quality
efficient for that type. The high type anticipates this quality reduction, and
the resulting intertemporal incentives discipline equilibrium behavior.

The comparison identifies the primitive source of the multiplicity in
\cref{thm:folk}. Quality differentiation alone is insufficient. What matters
is the combination of free disposal on the producer side and zero marginal
delivery cost on the broker side. Free disposal makes overprovision harmless
to the producer; zero marginal cost makes it harmless to the broker. Together,
they make efficient provision set-valued and generate the continuation
flexibility needed for equilibrium multiplicity.

\subsection{Relation to Coasian Bargaining}
\label{sec:coasian}

Our model relates to the literature on Coasian bargaining and durable-goods
monopoly. In the canonical setting, an uninformed seller repeatedly makes
price offers to a privately informed buyer. Because the seller cannot commit
to future prices, a buyer who rejects a current offer anticipates more
favorable terms in the future. In the gap case, where the lowest possible
buyer valuation is strictly positive, this force yields the familiar Coasian
prediction: as offers become arbitrarily frequent, trade occurs almost
immediately and the seller's profit converges to the lowest buyer valuation
\citep{FLT1985,gul1986foundations}.

Our simple-contract environment yields a different prediction. As the parties
become patient, the set of equilibrium broker payoffs does not collapse to a
point. Instead, \cref{thm:folk} implies that the set of equilibrium payoffs
contains an interval extending from $u_{\min}$ to arbitrarily close to the
commitment payoff $\pi(G)$. In this sense, our model delivers a
folk-theorem-type alternative to the standard Coasian selection result.

The closest precedent is \citet{ausubel1989reputation}, who show that the
Coasian conclusion can fail in the no-gap case. When the lowest buyer valuation
is zero, nonstationary reputational equilibria can support a broad range of
seller payoffs, including payoffs close to the static monopoly benchmark. Our
construction is analogous but relies on a different source of indifference. In
the no-gap environment of \citet{ausubel1989reputation}, the lowest buyer
derives zero surplus from the good, so delaying trade with that type is
costless. This indifference over the timing of trade creates the continuation
flexibility needed to support reputational equilibria.

In our model, by contrast, the lowest producer type has strictly positive
willingness to pay. The analogue of the Ausubel--Deneckere indifference arises
in the allocation dimension. Every feasible dataset contains the profiles that
the lowest type can profitably serve, so she is indifferent among all feasible
datasets. Thus, nonuniqueness of efficient dataset provision plays a role
analogous to nonuniqueness of efficient trade timing in the no-gap case.

There is, however, a further issue in relating our result to the Coase
conjecture. Once the object being sold has an endogenous quality dimension, the
natural analogue of a per-period offer is no longer obvious. One possibility is
to retain the canonical protocol by restricting the broker to a single
dataset--price offer in each period. Alternatively, one may view the
appropriate analogue for differentiated goods as allowing the seller to offer
a complete menu of qualities and prices in every period. Under this second
interpretation, the broker screens producer types intratemporally, implements
the commitment-optimal mechanism immediately (\cref{prop:menu}), and obtains
the commitment payoff for every $\delta<1$. Limited commitment then has no
effect on the equilibrium outcome. We therefore do not take a stand on whether
the simple-contract result should literally be viewed as a failure of the
Coase conjecture. Rather, we characterize how the predictions vary with the
contracting technology.

\section{Conclusion}
\label{sec:conclusion}

Data shares its cost structure with many modern goods: software, digital
content, and algorithmic services can all be replicated at negligible marginal
cost and freely discarded by their users. For such goods, efficient provision
need not be unique, and our analysis shows that this nonuniqueness has
strategic consequences: it can restore a seller's bargaining power under
limited commitment and make bargaining conventions a determinant of market
outcomes. Standard intuitions from durable-goods bargaining, developed for
goods with unique efficient allocations, need not carry over.

We have studied one producer, one broker, and a terminal sale. Competition
among brokers, resale, renewable access, and richer forms of ex-post moral
hazard all interact with the forces identified here in ways we do not yet
understand. We leave these questions to future work.

\bibliographystyle{ecta}
\bibliography{data} 
\newpage
\appendix
\section{Proofs}
\label{sec:proofs}

\subsection{Proofs of the Results in \cref{sec:commitment}}

\begin{proof}[Proof of \cref{prop:yang}]
Following standard arguments in static mechanism design, we first solve a relaxed problem in which only the local incentive constraints are imposed, and then verify that its solution satisfies the original incentive compatibility (IC) and individual rationality (IR) constraints.

Fix a mechanism $(q,\mu,p)$, where
$\mu_c\in\Delta(\mathcal B)$ is the dataset lottery assigned to type
$c$. For each report $c'$, let
\[
\overline B_{c'}
:=
\int_{\mathcal B}B\,\mathrm d\mu_{c'}(B)
\in L^1(F)
\]
be the barycenter of the assigned dataset lottery, and define the
composite allocation
\[
a_{c'}
:=
q(c')\,\overline B_{c'}
\in L^1(F).
\]
Then, for every true type $c$ and report $c'$, Fubini's theorem gives
\[
\begin{aligned}
q(c')\int_{\mathcal B}u(B,c)\,\mathrm d\mu_{c'}(B)
&=
q(c')
\int_{\underline v}^{\overline v}
\overline B_{c'}(v)(v-c)^+f(v)\,\mathrm dv\\
&=
\int_{\underline v}^{\overline v}
a_{c'}(v)(v-c)^+f(v)\,\mathrm dv.
\end{aligned}
\]
Thus $a_{c'}(v)\in[0,1]$ $F$-a.e.\ and, because every feasible
dataset equals one $F$-a.e.\ on
$[\overline c,\overline v]$,
\[
a_{c'}(v)=q(c')
\qquad
\text{for $F$-a.e. }v\ge\overline c.
\]
Every mechanism is therefore payoff-equivalent, for every producer
type and for the broker, to the composite allocation $a$ together with
the transfer rule $p$, and we work with $(a,p)$ from now on. Notice
that implementability additionally requires
$a_c(v)\le q(c)$ $F$-a.e. The relaxed problem below drops
this restriction away from $[\overline c,\overline v]$; this is
without loss because its optimizer has $q(c)=1$ and is
implementable by a deterministic dataset.

Define the indirect utility
\[
U(c):=\max_{c'\in[\underline{c},\overline{c}]}\left\{\int_{\underline{v}}^{\overline{v}} a_{c'}(v)\,(v-c)^+f(v)\,\mathrm{d}v-p(c')\right\}.
\]
For each fixed $c'$, the map $c\mapsto\int a_{c'}(v)(v-c)^+f(v)\,\mathrm{d}v$ is $1$-Lipschitz, since $(v-c)^+$ is $1$-Lipschitz in $c$ and $0\le a_{c'}f\le f$; by dominated convergence its derivative is $-\int a_{c'}(v)\mathbbm{1}_{\{v\ge c\}}f(v)\,\mathrm{d}v$, which is bounded in $[-1,0]$ uniformly in $c'$ and continuous in $c$ because $F$ is atomless. The envelope theorem therefore applies: under IC, $U$ is absolutely continuous and, for almost every $c$,
\begin{equation}
\label{eq:envelope}
U'(c)
=
-\int_{\underline v}^{\overline v}
a_c(v)\mathbbm 1_{\{v\ge c\}}f(v)\,\mathrm dv
\le0
\qquad
\text{for almost every }c.
\end{equation}
Thus $U$ is nonincreasing, so IR binds only at the highest cost type, and it is optimal to set $U(\overline{c})=0$. Then $U(c)=\int_c^{\overline{c}}\int a_s(v)\mathbbm{1}_{\{v\ge s\}}f(v)\,\mathrm{d}v\,\mathrm{d}s$, and the transfer of type $c$ is
\begin{equation}
\label{eq:transfer}
p(c)=\int_{\underline{v}}^{\overline{v}} a_c(v)(v-c)^+f(v)\,\mathrm{d}v-U(c).
\end{equation}
Substituting \eqref{eq:transfer} into the broker's objective and exchanging the order of integration in $\int_{\underline{c}}^{\overline{c}}U(c)g(c)\,\mathrm{d}c$ (Fubini; the integrand is bounded) yields
\[
\int_{\underline{c}}^{\overline{c}} p(c)\,g(c)\,\mathrm{d}c
=\int_{\underline{c}}^{\overline{c}}\left[\int_{\underline{v}}^{\overline{v}} a_c(v)\,\mathbbm{1}_{\{v\ge c\}}\Big(v-c-\frac{G(c)}{g(c)}\Big)f(v)\,\mathrm{d}v\right]g(c)\,\mathrm{d}c .
\]
The relaxed problem is to maximize this expression over measurable $a$ with $a_c(v)\in[0,1]$ and $a_c(v)=q(c)$ on $[\overline{c},\overline{v}]$. The objective is linear in $a_c(v)$ pointwise, so an optimal solution is bang-bang: since $\phi(c)=c+G(c)/g(c)\ge c$, for $v<\overline{c}$ it is optimal to set $a_c(v)=1$ if $v\ge\phi(c)$ and $a_c(v)=0$ otherwise. On $[\overline{c},\overline{v}]$ the allocation must equal $q(c)$, and $q(c)=1$ is optimal provided
\[
\int_{\overline{c}}^{\overline{v}}\big(v-\phi(c)\big)f(v)\,\mathrm{d}v\ \ge 0 .
\]
This holds trivially when $\phi(c)\le\overline{c}$. When $\phi(c)>\overline{c}$, monotonicity of $\phi$ gives $\phi(c)\le\phi(\overline{c})=\overline{c}+1/g(\overline{c})$, so the integral is at least $\int_{\overline{c}}^{\overline{v}}\big(v-\overline{c}-1/g(\overline{c})\big)f(v)\,\mathrm{d}v\ge0$ by \cref{ass:no-exclusion}. Hence the relaxed problem is solved by $q(c)=1$ and the deterministic dataset
\[
B(c)=\Big[\min\Big\{c+\frac{G(c)}{g(c)},\,\overline{c}\Big\},\,\overline{v}\Big],
\]
that is, $a_c(v)=\mathbbm{1}\{v\ge\min\{\phi(c),\overline{c}\}\}$. In particular, randomization over datasets and exclusion are never strictly beneficial.

The utility function exhibits single crossing in the almost-everywhere
inclusion order. If $B\supseteq B'$, then
\[
u(B,c)-u(B',c)
=
\int_{\underline v}^{\overline v}
\bigl(B(v)-B'(v)\bigr)(v-c)^+f(v)\,\mathrm dv,
\]
which is nonincreasing in $c$. Hence lower-cost types value
additional data weakly more than higher-cost types. Because $\min\{\phi(c),\overline{c}\}$ is nondecreasing in $c$, the allocation $B(c)$ is nonincreasing in $c$ under set inclusion, which is the monotonicity condition for this order. Together with transfers given by \eqref{eq:transfer}, monotonicity and the envelope condition \eqref{eq:envelope} imply global incentive compatibility by the standard argument. Individual rationality follows from $U(\overline{c})=0$ and $U$ nonincreasing. The candidate is therefore feasible for the original problem and optimal for the relaxed one, hence optimal. Its value is $\pi(G)$.
\end{proof}

\begin{proof}[Proof of \cref{cor:yang}]
Fix a broker strategy $\sigma$ and a producer best response $\alpha$.
For each producer type $c$ and period $t\in\mathbb N$, let
$\eta_c^t$ be the finite measure on $\mathcal B\times P$ defined by
\[
\eta_c^t(E)
:=
\Pr_c\bigl((B_T,p_T)\in E,\ T=t\bigr),
\qquad
E\in\mathscr B(\mathcal B\times P),
\]
where $T$ is the period in which acceptance occurs and
$(B_T,p_T)$ is the accepted contract. Thus
\[
\sum_{t=0}^{\infty}
\eta_c^t(\mathcal B\times P)
\le1,
\]
with the missing mass corresponding to no trade. Because $P$ is
compact, all discounted transfers below are absolutely integrable.

Define a static mechanism $(q,\mu,p)$ by
\[
q(c)
:=
\sum_{t=0}^{\infty}
\delta^t\eta_c^t(\mathcal B\times P)
\in[0,1],
\]
and, when $q(c)>0$, define the dataset lottery $\mu_c$ by
\[
\mu_c(E)
:=
\frac{1}{q(c)}
\sum_{t=0}^{\infty}
\delta^t\eta_c^t(E\times P),
\qquad
E\in\mathscr B(\mathcal B).
\]
When $q(c)=0$, let $\mu_c$ be arbitrary. Finally, define the expected
transfer by
\[
p(c)
:=
\sum_{t=0}^{\infty}
\delta^t
\int_{\mathcal B\times P}
r\,\mathrm d\eta_c^t(B,r).
\]
Standard measurability of the outcome laws induced by Borel stochastic
kernels implies that $c\mapsto(q(c),\mu_c,p(c))$ is a measurable
mechanism.

For any true type $c$ and any reported type $c'$, the payoff generated
by the static contract assigned to $c'$ is
\[
\begin{aligned}
q(c')
\int_{\mathcal B}u(B,c)\,\mathrm d\mu_{c'}(B)
-p(c')
&=
\sum_{t=0}^{\infty}
\delta^t
\int_{\mathcal B\times P}
\bigl[u(B,c)-r\bigr]\,
\mathrm d\eta_{c'}^t(B,r).
\end{aligned}
\]
The right-hand side is exactly the expected payoff of a true type $c$
who, in the dynamic game, follows the entire behavioral strategy
prescribed by $\alpha$ for type $c'$. Because the broker never
observes the producer's true type, this deviation induces the same
distribution $\{\eta_{c'}^t\}_{t\ge0}$ as it does for type $c'$.

Since $\alpha$ is a best response, truthful behavior weakly dominates
this deviation for every pair $(c,c')$. The induced static mechanism
therefore satisfies incentive compatibility. It also satisfies
individual rationality, because rejecting every offer in the dynamic
game is feasible and yields zero.

By construction,
\[
\int_{\underline c}^{\overline c}
p(c)g(c)\,\mathrm dc
\]
equals the broker's expected discounted payoff under
$(\sigma,\alpha)$. Hence every payoff generated by a broker strategy
and a producer best response in the dynamic game is generated by an
incentive-compatible and individually rational static mechanism.
The broker's dynamic payoff can therefore never exceed the
commitment payoff $\pi(G)$.
\end{proof}

\subsection{Proofs of the Results in \cref{sec:coase}}
The following two auxiliary lemmas establish that immediate market clearing is always feasible for the broker. Recall from Section~\ref{sec:model} that $A(h)\subseteq[\underline{c},\overline{c}]$ denotes the set of active producer types at $h$, that is, those who have not yet purchased, and $\overline{c}(h):=\sup A(h)$ denotes its highest-cost element, that is, the active type with the lowest willingness to pay. Two properties of $u(B,c)=\int_B (v-c)^+f(v)\,\mathrm{d}v$ are used repeatedly. For every feasible $B$ and $c\le c'$,
\begin{equation}
\label{eq:lipschitz}
(c'-c)\,\big(1-F(\overline{c})\big)\ \le\ u(B,c)-u(B,c')\ \le\ c'-c,
\end{equation}
where the lower bound uses $[\overline{c},\overline{v}]\subseteq B$ and $F(\overline{c})<1$. In particular, $u(B,\cdot)$ is strictly decreasing, so lower-cost types can mimic higher-cost types and obtain strictly higher utility.

\begin{lemma}
\label{lem:nopositive}
Let $(\sigma,\alpha)$ be an equilibrium. For any history $h_t$ and any offer $(B_t,p_t)$ in the support of $\sigma(h_t)$,
\[
u\big(B_t,\overline{c}(h_t)\big)-p_t\ \le\ 0 .
\]
\end{lemma}

\begin{proof}[Proof of \cref{lem:nopositive}]
Define
\[
d_{\sup}:=\sup_{h\in H}\ \sup_{(B,p)\in\operatorname{supp}\sigma(h)}\Big\{u\big(B,\overline{c}(h)\big)-p\Big\}.
\]
Suppose, for a contradiction, that $d_{\sup}>0$. Fix $\epsilon>0$ small enough that
\begin{equation}
\label{eq:eps}
d_{\sup}-2\epsilon\ >\ \delta\,(d_{\sup}+\epsilon),
\end{equation}
and choose a history $h^*$ and an offer $(B^*,p^*)\in\operatorname{supp}\sigma(h^*)$ with $u(B^*,\overline{c}(h^*))-p^*>d_{\sup}-\epsilon$.

\smallskip
\noindent\textit{Claim: every active type at $h^*$ accepts any offer $(B^*,p)$ with $p\le p^*+\epsilon$.}
Accepting yields every type $c\in A(h^*)$ at least
\[
u(B^*,c)-p\ \ge\ u\big(B^*,\overline{c}(h^*)\big)-(p^*+\epsilon)\ >\ d_{\sup}-2\epsilon ,
\]
by monotonicity of $u$ in $c$. Suppose instead that a nonempty set of types rejects, and let $h'$ denote the resulting history, so that $A(h')\ne\emptyset$. By definition of $d_{\sup}$, at any history $h\succeq h'$ every offer $(B,p)$ in the support of $\sigma(h)$ satisfies $p\ge u(B,\overline{c}(h))-d_{\sup}$. Since $A(h)\subseteq A(h')$ implies $\overline{c}(h)\le\overline{c}(h')$, a type $c\in A(h')$ that accepts $(B,p)$ at such an $h$ obtains, by \eqref{eq:lipschitz},
\[
u(B,c)-p\ \le\ u(B,c)-u\big(B,\overline{c}(h)\big)+d_{\sup}\ \le\ \big(\overline{c}(h')-c\big)+d_{\sup}.
\]
Pick $c\in A(h')$ with $\overline{c}(h')-c<\epsilon$, which exists by definition of the supremum. Any strategy that rejects at $h^*$ yields this type at most $\delta(d_{\sup}+\epsilon)$, whereas accepting yields more than $d_{\sup}-2\epsilon$. By \eqref{eq:eps}, rejecting is not a best response, contradicting the optimality of $\alpha$. This proves the claim.

\smallskip
By the claim applied to $p=p^*$, the offer $(B^*,p^*)$ clears the market at $h^*$ and yields the broker $p^*$. By the claim applied to $p=p^*+\epsilon$, the offer $(B^*,p^*+\epsilon)$ also clears the market and yields $p^*+\epsilon$, which is strictly higher. Since $(B^*,p^*)$ lies in the support of $\sigma(h^*)$, it must be optimal at $h^*$, a contradiction. Hence $d_{\sup}\le 0$, which is the statement of the lemma.
\end{proof}

\begin{lemma}
\label{lem:clear}
Let $(\sigma,\alpha)$ be an equilibrium and $h_t$ a history. Any offer $(B_t,p_t)$ satisfying either
\begin{enumerate}
\item[(i)] $u(B_t,\overline{c}(h_t))-p_t>0$, or
\item[(ii)] $u(B_t,\overline{c}(h_t))-p_t=0$ and $[\overline{c}(h_t),\overline{v}]\subseteq B_t$,
\end{enumerate}
clears the market immediately, provided that type $\overline{c}(h_t)$ breaks indifference in favor of trade. Condition (ii) is automatically satisfied whenever $\overline{c}(h_t)=\overline{c}$, since every feasible dataset contains $[\overline{c},\overline{v}]$.
\end{lemma}

\begin{proof}[Proof of \cref{lem:clear}]
Write $c^*:=\overline{c}(h_t)$, and suppose, for a contradiction, that a nonempty set of types rejects $(B_t,p_t)$; let $h_{t+1}$ denote the resulting history, so $A(h_{t+1})\ne\emptyset$ and $\overline{c}(h_{t+1})\le c^*$.

We first bound the payoff from waiting. At any history $h\succeq h_{t+1}$, \cref{lem:nopositive} gives $p\ge u(B,\overline{c}(h))$ for every offer $(B,p)$ in the support of $\sigma(h)$, and $\overline{c}(h)\le\overline{c}(h_{t+1})$. Hence a type $c\in A(h_{t+1})$ that accepts $(B,p)$ at $h$ obtains, using monotonicity of $u$ in $c$ and \eqref{eq:lipschitz},
\begin{equation}
\label{eq:waitbound}
u(B,c)-p\ \le\ u(B,c)-u\big(B,\overline{c}(h_{t+1})\big)\ \le\ \overline{c}(h_{t+1})-c .
\end{equation}
Since acceptance occurs no earlier than period $t+1$, the continuation payoff of type $c\in A(h_{t+1})$ from rejecting at $h_t$ is at most $\delta\,\big(\overline{c}(h_{t+1})-c\big)$.

\smallskip
\noindent\textit{Case (i).}
For $c\in A(h_{t+1})$, purchasing at $h_t$ yields $u(B_t,c)-p_t\ge u(B_t,c^*)-p_t>0$, a bound independent of $c$. Choosing $c\in A(h_{t+1})$ with $\overline{c}(h_{t+1})-c$ sufficiently small, the waiting payoff $\delta(\overline{c}(h_{t+1})-c)$ falls below this bound, so type $c$ strictly prefers to purchase at $h_t$, contradicting sequential rationality.

\smallskip
\noindent\textit{Case (ii).}
If $\overline{c}(h_{t+1})<c^*$, then for $c\in A(h_{t+1})$ the immediate payoff satisfies, by \eqref{eq:lipschitz},
\[
u(B_t,c)-p_t\ =\ u(B_t,c)-u(B_t,c^*)\ \ge\ \big(c^*-\overline{c}(h_{t+1})\big)\big(1-F(\overline{c})\big)\ >\ 0,
\]
again a bound independent of $c$, and the argument of Case (i) applies verbatim.

If $\overline{c}(h_{t+1})=c^*$, take $c\in A(h_{t+1})$ with $c<c^*$. Refining \eqref{eq:waitbound} by computing the supremum over datasets explicitly, the waiting payoff is at most
\begin{align*}
\delta\,\sup_{B\in\mathcal{B}}\big[u(B,c)-u(B,c^*)\big]
&=\delta\left[\int_{c}^{c^*}(v-c)f(v)\,\mathrm{d}v+(c^*-c)\big(1-F(c^*)\big)\right]\\
&\le\delta\,(c^*-c)\Big[\big(F(c^*)-F(c)\big)+\big(1-F(c^*)\big)\Big],
\end{align*}
where the supremum is attained at $B=[\underline{v},\overline{v}]$ because the integrand $(v-c)^+-(v-c^*)^+$ is nonnegative. On the other hand, since $p_t=u(B_t,c^*)$ and $[c^*,\overline{v}]\subseteq B_t$, purchasing immediately yields
\[
u(B_t,c)-p_t
=\int_{B_t}\big[(v-c)^+-(v-c^*)^+\big]f(v)\,\mathrm{d}v
\ \ge\ \int_{c^*}^{\overline{v}}(c^*-c)f(v)\,\mathrm{d}v
=(c^*-c)\big(1-F(c^*)\big).
\]
As $c\uparrow c^*$, $F(c^*)-F(c)\to0$ while $1-F(c^*)\ge 1-F(\overline{c})>0$ is fixed, so for $\delta<1$ and $c$ sufficiently close to $c^*$,
\[
\delta\,(c^*-c)\Big[\big(F(c^*)-F(c)\big)+\big(1-F(c^*)\big)\Big]\ <\ (c^*-c)\big(1-F(c^*)\big).
\]
Type $c$ therefore strictly prefers to purchase at $h_t$, contradicting sequential rationality. Finally, type $c^*$ itself is indifferent and, by assumption, accepts. Hence the market clears at $h_t$.
\end{proof}

\begin{proof}[Proof of \cref{thm:Coase}]
We begin with a preliminary lemma that characterizes the optimal uniform price when all remaining types receive the full dataset.

\begin{lemma}
\label{lem:use}
Fix $c'\in[\underline{c},\overline{c}]$ and let $G_{c'}(\cdot)$ denote the producer type distribution conditional on $[c',\overline{c}]$. Suppose the broker offers the full dataset $[\underline{v},\overline{v}]$ at a single uniform price $p$ to these types. Then the revenue-maximizing price is $p=u_{\min}$. More precisely, the revenue function
\[
R_{c'}(c):=u\big([\underline{v},\overline{v}],c\big)\,\big(G(c)-G(c')\big),\qquad c\in[c',\overline{c}],
\]
obtained by charging the price that makes type $c$ indifferent, is nondecreasing in $c$.
\end{lemma}

\begin{proof}[Proof of \cref{lem:use}]
Since $u([\underline{v},\overline{v}],\cdot)$ is strictly decreasing, a uniform price $p\in[u_{\min},u([\underline{v},\overline{v}],c')]$ is accepted exactly by the types $c$ with $u([\underline{v},\overline{v}],c)\ge p$, that is, by $[c',c(p)]$ where $u([\underline{v},\overline{v}],c(p))=p$. Revenue is therefore $R_{c'}(c(p))$, and prices outside this interval are dominated. Differentiating, and using $\partial_c u([\underline{v},\overline{v}],c)=-(1-F(c))$,
\begin{align*}
R_{c'}'(c)
= g(c)\left[u\big([\underline{v},\overline{v}],c\big)-\frac{G(c)-G(c')}{g(c)}\big(1-F(c)\big)\right]
\ &\ge\ g(c)\left[u\big([\underline{v},\overline{v}],c\big)-\frac{G(c)}{g(c)}\big(1-F(c)\big)\right]\\
&= g(c)\int_{c}^{\overline{v}}\Big(v-c-\frac{G(c)}{g(c)}\Big)f(v)\,\mathrm{d}v
\ \ge\ 0,
\end{align*}
where the first inequality uses $G(c)-G(c')\le G(c)$ and the last is \cref{ass:no-exclusion}. Hence $R_{c'}$ is maximized at $c=\overline{c}$, that is, at the market-clearing price $u([\underline{v},\overline{v}],\overline{c})=u_{\min}$.
\end{proof}

\medskip
\noindent
We now turn to the proof of the proposition. Throughout, the continuation play after any offer $(B_t,p_t)$ is that the broker offers $(B_t,u_{\min})$ in period $t+1$, which clears the market by \cref{lem:clear}(ii) since $u(B_t,\overline{c})=u_{\min}$ for every feasible $B_t$. Given this continuation, an active type $c$ weakly prefers to accept $(B_t,p_t)$ immediately if and only if
\begin{equation}
\label{eq:accept}
u(B_t,c)-p_t\ \ge\ \delta\big(u(B_t,c)-u_{\min}\big)
\quad\Longleftrightarrow\quad
(1-\delta)\,u(B_t,c)\ \ge\ p_t-\delta u_{\min}.
\end{equation}
Since $u(B_t,\cdot)$ is continuous and strictly decreasing, the set of types satisfying \eqref{eq:accept} is a lower interval $[\underline{c},c(\delta,B_t,p_t)]$, where the cutoff is defined as follows:
\begin{enumerate}
\item If $p_t\le u_{\min}$, set $c(\delta,B_t,p_t)=\overline{c}$: every active type accepts.
\item If $u_{\min}<p_t\le(1-\delta)u(B_t,\underline{c})+\delta u_{\min}$, let $c(\delta,B_t,p_t)\in[\underline{c},\overline{c})$ be the unique solution of $(1-\delta)u(B_t,c)=p_t-\delta u_{\min}$: types $c\le c(\delta,B_t,p_t)$ accept and higher-cost types wait.
\item If $p_t>(1-\delta)u(B_t,\underline{c})+\delta u_{\min}$, the acceptance set is empty and every active type waits.
\end{enumerate}
In particular, although the skimming property may fail in general, the persistence of the dataset across periods ensures that under the strategies below the acceptance set at every history is an interval of the lowest-cost active types, so the active set at any history is of the form $[c',\overline{c}]$ (up to null sets).

\medskip
\noindent
We consider the following candidate equilibrium strategies:
\begin{enumerate}
\item \emph{Broker:} In each period $t$, offer $(B_{t-1},u_{\min})$, where $B_{t-1}$ is the dataset offered in the previous period, starting from $B_{-1}=[\underline{v},\overline{v}]$. Denote this strategy by $\sigma$.
\item \emph{Producer:} In period $t$, upon observing $(B_t,p_t)$, accept if and only if \eqref{eq:accept} holds, with indifference resolved in favor of acceptance. Denote this strategy by $\alpha$.
\end{enumerate}
On the equilibrium path, the broker offers $([\underline{v},\overline{v}],u_{\min})$ at $t=0$, every type accepts, and the broker's revenue is $u_{\min}$.

\medskip
\noindent
\textit{Producer optimality.} At any history, after any offer $(B_t,p_t)$, the producer anticipates that $\sigma$ prescribes $(B_t,u_{\min})$ in period $t+1$ and, should she reject again, $(B_t,u_{\min})$ in every subsequent period. Accepting in period $t+1$ yields $\delta(u(B_t,c)-u_{\min})\ge0$, which weakly dominates accepting later or never. Hence the best response is exactly \eqref{eq:accept}, and $\alpha$ is sequentially rational.

\medskip
\noindent
\textit{Broker optimality.} Since $\delta<1$ and payoffs are bounded, it suffices by the one-shot deviation principle to check single-period deviations. Fix a history $h_t$ with active set $[c',\overline{c}]$ and previous dataset $B_{t-1}$. The prescribed offer $(B_{t-1},u_{\min})$ clears the market immediately (case 1 above) and yields
\[
u_{\min}\,\big(1-G(c')\big).
\]
Consider a one-shot deviation to $(B,p)$ followed by reversion to $\sigma$, so that the residual market is cleared at $(B,u_{\min})$ in period $t+1$.
\begin{itemize}
\item If $p\le u_{\min}$, the market clears immediately at a weakly lower price, so the deviation is weakly dominated.
\item If the acceptance set is empty (case 3), the deviation yields $\delta u_{\min}(1-G(c'))<u_{\min}(1-G(c'))$.
\item Otherwise (case 2), let $c^*=c(\delta,B,p)\in[c',\overline{c})$, so that $p=(1-\delta)u(B,c^*)+\delta u_{\min}$. The deviation yields
\[
p\,\big(G(c^*)-G(c')\big)+\delta u_{\min}\big(1-G(c^*)\big),
\]
and the gain relative to immediate clearing is
\begin{align*}
&\big(p-u_{\min}\big)\big(G(c^*)-G(c')\big)-(1-\delta)\,u_{\min}\big(1-G(c^*)\big)\\
&\qquad=(1-\delta)\Big[u(B,c^*)\big(G(c^*)-G(c')\big)-u_{\min}\big(1-G(c')\big)\Big]\\
&\qquad\le(1-\delta)\Big[R_{c'}(c^*)-R_{c'}(\overline{c})\Big]\ \le\ 0,
\end{align*}
where the first inequality uses $u(B,c^*)\le u([\underline{v},\overline{v}],c^*)$ and $R_{c'}(\overline{c})=u_{\min}(1-G(c'))$, and the second is \cref{lem:use}. If $c^*<\overline{c}$ and $G(c^*)<1$ the last inequality can be made strict for the full dataset whenever $R_{c'}$ is strictly increasing, but weak dominance suffices.
\end{itemize}
No one-shot deviation is profitable at any history, so $\sigma$ is sequentially rational. We conclude that $(\sigma,\alpha)$ constitutes an equilibrium for all $0\le\delta<1$, with the broker's payoff equal to $u_{\min}$.
\end{proof}

\begin{proof}[Proof of \cref{thm:folk}]
\noindent We proceed in two steps. Throughout, $B^*(c)=[\min\{\phi(c),\overline{c}\},\overline{v}]$ denotes the commitment-optimal allocation of \cref{prop:yang}, and $c^*\in(\underline{c},\overline{c})$ denotes the unique type with $\phi(c^*)=\overline{c}$, which exists since $\phi(\underline{c})=\underline{c}<\overline{c}<\phi(\overline{c})$ and $\phi$ is continuous and strictly increasing. Types $c\ge c^*$ receive $B^*(c)=[\overline{c},\overline{v}]$ and pay $u_{\min}$ in the commitment-optimal mechanism. Recall from the proof of \cref{prop:yang} that for any incentive-compatible mechanism with composite allocation $a$ and $U(\overline{c})=0$, revenue equals
\begin{equation}
\label{eq:revenue}
\int_{\underline{c}}^{\overline{c}}\left[\int_{\underline{v}}^{\overline{v}} a_c(v)\,\mathbbm{1}_{\{v\ge c\}}\big(v-\phi(c)\big)f(v)\,\mathrm{d}v\right]g(c)\,\mathrm{d}c,
\end{equation}
and that $\pi(G)$ is the value of \eqref{eq:revenue} at $a_c=\mathbbm{1}_{B^*(c)}$.

\begin{enumerate}[label=\textbf{Step \arabic*.}, wide=0pt, leftmargin=*, align=left, labelwidth=*, itemsep=1ex, listparindent=\parindent, parsep=0pt]

\item \textbf{Construction of the high-payoff equilibrium.}
We show that for any $\epsilon>0$ there exists $\underline{\delta}<1$ such that for all $\delta\ge\underline{\delta}$ there is an equilibrium with broker payoff at least $\pi(G)-\epsilon$.

Fix $n\ge2$ and a partition
\[
\underline{c}=c_0<c_1<\dots<c_{n-1}<c_n=c^*
\]
of $[\underline{c},c^*]$ with mesh $\max_i|c_i-c_{i-1}|\to0$ as $n\to\infty$. Assign datasets
\[
B_i=\big[\phi(c_i),\overline{v}\big]\ \text{ for } i=1,\dots,n-1,
\qquad
B_n=\big[\overline{c},\overline{v}\big],
\]
so that $B_1\supsetneq B_2\supsetneq\dots\supsetneq B_n$, with each difference $B_i\setminus B_{i+1}=[\phi(c_i),\phi(c_{i+1}))$ of strictly positive $F$-measure. Set $p_n=u_{\min}$ and define $p_{n-1},\dots,p_1$ recursively by the indifference conditions
\begin{equation}
\label{eq:indiff}
u(B_i,c_i)-p_i=\delta\,\big[u(B_{i+1},c_i)-p_{i+1}\big],\qquad i=1,\dots,n-1.
\end{equation}
Let $m_i:=G(c_i)-G(c_{i-1})$ for $i\le n-1$ and $m_n:=1-G(c_{n-1})$ denote the masses of the groups $[c_{i-1},c_i]$ and $[c_{n-1},\overline{c}]$.

Consider the following strategy profile.
\begin{enumerate}
\item \emph{On path:} In period $t=i-1$, $i=1,\dots,n-1$, the broker posts $(B_i,p_i)$ and all active types $c\in[c_{i-1},c_i]$ accept. In period $t=n-1$ the broker posts $(B_n,p_n)=([\overline{c},\overline{v}],u_{\min})$ and all remaining types $[c_{n-1},\overline{c}]$ accept.
\item \emph{Off path:} Any deviation by the broker triggers an immediate and permanent reversion to the strategies $(\sigma,\alpha)$ of \cref{thm:Coase}, starting with the producer's response to the deviating offer itself.
\end{enumerate}

On the equilibrium path a type $c$ chooses a stage $j\in\{1,\dots,n\}$ at which to accept (or never accepts, which yields $0$), obtaining $W_j(c):=\delta^{j-1}\big[u(B_j,c)-p_j\big]$. Using \eqref{eq:indiff},
\[
W_i(c)-W_{i+1}(c)=\delta^{i-1}\Big\{\big[u(B_i,c)-\delta u(B_{i+1},c)\big]-\big[u(B_i,c_i)-\delta u(B_{i+1},c_i)\big]\Big\}.
\]
The map $c\mapsto u(B_i,c)-\delta u(B_{i+1},c)$ has derivative $-\int_{B_i}\mathbbm{1}_{\{v\ge c\}}f\,\mathrm{d}v+\delta\int_{B_{i+1}}\mathbbm{1}_{\{v\ge c\}}f\,\mathrm{d}v\le-(1-\delta)\big(1-F(\overline{c})\big)<0$, since $B_i\supseteq B_{i+1}\supseteq[\overline{c},\overline{v}]$. Hence $W_i(c)\ge W_{i+1}(c)$ if and only if $c\le c_i$. For $c\in[c_{i-1},c_i]$ this gives $W_i(c)\ge W_{i+1}(c)\ge\dots\ge W_n(c)$ (as $c\le c_i\le c_{i+1}\le\cdots$) and $W_i(c)\ge W_{i-1}(c)\ge\dots\ge W_1(c)$ (as $c\ge c_{i-1}\ge c_{i-2}\ge\cdots$). Finally $W_n(c)=\delta^{n-1}[u([\overline{c},\overline{v}],c)-u_{\min}]\ge0$, so accepting at stage $i$ is optimal for every $c\in[c_{i-1},c_i]$, and likewise at stage $n$ for $c\in[c_{n-1},\overline{c}]$. Off path, $\alpha$ is a best response by \cref{thm:Coase}.

By the one-shot deviation principle it suffices to consider a single deviation at an on-path history, that is, at stage $i$ with active set $[c_{i-1},\overline{c}]$ of mass $1-G(c_{i-1})$. Following the deviation, play is governed by $(\sigma,\alpha)$ of \cref{thm:Coase}; the broker-optimality step of that proof shows that any offer $(B,p)$ followed by this continuation yields at most $u_{\min}\big(1-G(c_{i-1})\big)$. The on-path continuation payoff at stage $i$ is
\[
\Pi_i(\delta):=\sum_{j=i}^{n}\delta^{\,j-i}\,m_j\,p_j,
\]
where the $p_j$ depend on $\delta$ through \eqref{eq:indiff}. We claim that for fixed $n$ there is $\underline{\delta}_1(n)<1$ such that $\Pi_i(\delta)\ge u_{\min}(1-G(c_{i-1}))$ for all $i$ and all $\delta\ge\underline{\delta}_1(n)$. Indeed, each $p_j$ is a polynomial in $\delta$, and at $\delta=1$ the recursion gives
\[
p_j(1)=u_{\min}+\sum_{k=j}^{n-1}\big[u(B_k,c_k)-u(B_{k+1},c_k)\big]
=u_{\min}+\sum_{k=j}^{n-1}\int_{\phi(c_k)}^{\phi(c_{k+1})}(v-c_k)f(v)\,\mathrm{d}v\ >\ u_{\min}
\]
for $j\le n-1$ (with $\phi(c_n)=\overline{c}$), because each integrand is positive on a set of positive measure. Since $m_j>0$, it follows that $\Pi_i(1)=\sum_{j\ge i}m_jp_j(1)>u_{\min}(1-G(c_{i-1}))$ for $i\le n-1$, with equality at $i=n$. By continuity in $\delta$ the claim follows. Hence for $\delta\ge\underline{\delta}_1(n)$ no deviation is profitable and the profile is an equilibrium.

Define the step allocation $B_n(c):=B_i$ for $c\in(c_{i-1},c_i]$, $i\le n-1$, and $B_n(c):=B_n$ for $c\in(c_{n-1},\overline{c}]$, with $t(c):=i-1$ on the corresponding intervals. The equilibrium outcome is incentive compatible with $U(\overline{c})=\delta^{n-1}[u([\overline{c},\overline{v}],\overline{c})-u_{\min}]=0$, so by \cref{cor:yang} and \eqref{eq:revenue} with $a_c(v)=\delta^{t(c)}\mathbbm{1}_{B_n(c)}(v)$, the broker's payoff is
\[
\Pi_1(\delta)=\sum_{i=1}^n\delta^{i-1}m_ip_i
=\int_{\underline{c}}^{\overline{c}}\delta^{t(c)}\left[\int_{B_n(c)}\mathbbm{1}_{\{v\ge c\}}\big(v-\phi(c)\big)f(v)\,\mathrm{d}v\right]g(c)\,\mathrm{d}c .
\]
Denote by $\Psi_n$ the same integral without the factor $\delta^{t(c)}$.

As $n\to\infty$, $B_n(c)\to B^*(c)$ for every $c$: on $(c_{i-1},c_i]$ we have $\phi(c_i)\to\phi(c)$ by continuity of $\phi$ and $c_i-c\le$ mesh, so $\mathbbm{1}_{B_n(c)}\to\mathbbm{1}_{B^*(c)}$ a.e.\ in $v$, while on $[c^*,\overline{c}]$ the two coincide. The inner integrand is bounded by $\overline{v}M$, so by dominated convergence $\Psi_n\to\pi(G)$, and there is $N$ with $|\pi(G)-\Psi_n|\le\epsilon/2$ for all $n\ge N$. Fix such an $n$. Since $t(c)\le n-1$, $\delta^{t(c)}\to1$ uniformly as $\delta\to1$, so there is $\underline{\delta}_2(n)<1$ with $|\Pi_1(\delta)-\Psi_n|\le\epsilon/2$ for all $\delta\ge\underline{\delta}_2(n)$. Setting $\underline{\delta}:=\max\{\underline{\delta}_1(n),\underline{\delta}_2(n)\}$ and using the triangle inequality, the equilibrium payoff satisfies $\Pi_1(\delta)\ge\pi(G)-\epsilon$ for all $\delta\ge\underline{\delta}$.

\item \textbf{Filling the continuum.}
Fix $n$ and $\delta\ge\underline\delta$ as in Step~1, increasing
$\underline\delta$ if necessary so that $\underline\delta>0$. Denote
the cutoffs, datasets, prices, group masses, and continuation payoffs
of the Step~1 equilibrium by
\[
(c_i)_{i=0}^n,\qquad
(B_i^0)_{i=1}^n,\qquad
(p_i^0)_{i=1}^n,\qquad
(m_i^0)_{i=1}^n,\qquad
(\Pi_i^0)_{i=1}^n,
\]
where $B_i^0=B_i$, $p_i^0=p_i$, $m_i^0=m_i$, and
$\Pi_i^0=\Pi_i(\delta)$. In particular,
\[
\Pi^{\mathrm{high}}
:=
\Pi_1^0
\ge
\pi(G)-\epsilon
>
u_{\min}.
\]

For any weakly increasing cutoff vector
\[
\widetilde{\mathbf c}
=
(\widetilde c_1,\ldots,\widetilde c_{n-1})
\quad\text{with}\quad
\underline c
\le
\widetilde c_1
\le\cdots\le
\widetilde c_{n-1}
\le c^*,
\]
set $\widetilde c_0:=\underline c$ and define
\[
\widetilde B_i:=B^*(\widetilde c_i),
\qquad i=1,\ldots,n-1,
\qquad
\widetilde B_n:=[\overline c,\overline v].
\]
Set $\widetilde p_n:=u_{\min}$ and recursively define
$\widetilde p_{n-1},\ldots,\widetilde p_1$ from
\begin{equation}
\label{eq:indiff-general}
u(\widetilde B_i,\widetilde c_i)-\widetilde p_i
=
\delta
\bigl[
u(\widetilde B_{i+1},\widetilde c_i)
-\widetilde p_{i+1}
\bigr],
\qquad
i=1,\ldots,n-1.
\end{equation}
The corresponding group masses are
\[
\widetilde m_i
:=
G(\widetilde c_i)-G(\widetilde c_{i-1}),
\qquad
i=1,\ldots,n-1,
\]
and
\[
\widetilde m_n
:=
1-G(\widetilde c_{n-1}).
\]
For every stage $j$, define the broker's continuation payoff by
\[
\widetilde\Pi_j(\widetilde{\mathbf c})
:=
\sum_{k=j}^{n}
\delta^{\,k-j}
\widetilde m_k\widetilde p_k.
\]

The associated strategy profile is the one from Step~1: at stage $i$,
the broker offers $(\widetilde B_i,\widetilde p_i)$, types in
$[\widetilde c_{i-1},\widetilde c_i]$ accept, and any broker deviation
triggers the punishment equilibrium of
\cref{thm:Coase}. Consecutive cutoffs may coincide, in
which case the corresponding group has zero mass.

The producer incentive argument from Step~1 continues to apply. To
see this directly, define
\[
\widetilde W_j(c)
:=
\delta^{\,j-1}
\bigl[u(\widetilde B_j,c)-\widetilde p_j\bigr].
\]
Equation \eqref{eq:indiff-general} implies that
$\widetilde W_i(\widetilde c_i)
=\widetilde W_{i+1}(\widetilde c_i)$. Moreover,
\[
\begin{aligned}
\frac{\partial}{\partial c}
\left[
u(\widetilde B_i,c)
-\delta u(\widetilde B_{i+1},c)
\right]
&=
-\int_{\underline v}^{\overline v}
\left[
\widetilde B_i(v)
-\delta\widetilde B_{i+1}(v)
\right]
\mathbbm 1_{\{v\ge c\}}
f(v)\,\mathrm dv\\
&\le
-(1-\delta)\bigl(1-F(\overline c)\bigr)
<0.
\end{aligned}
\]
Hence
\[
\widetilde W_i(c)\ge\widetilde W_{i+1}(c)
\quad\Longleftrightarrow\quad
c\le\widetilde c_i.
\]
It follows by adjacent comparisons that every type optimally accepts
at its assigned stage. Finally,
\[
\widetilde W_n(c)
=
\delta^{n-1}
\bigl[
u([\overline c,\overline v],c)-u_{\min}
\bigr]
\ge0,
\]
so accepting at the assigned stage also weakly dominates never
trading.

We record one property of the recursive prices that will be used
below:
\begin{equation}
\label{eq:prices-above-umin}
\widetilde p_i\ge u_{\min},
\qquad
i=1,\ldots,n.
\end{equation}
This follows by backward induction. The claim holds for $i=n$. If it
holds for $i+1$, then \eqref{eq:indiff-general} gives
\begin{align*}
\widetilde p_i-u_{\min}
={}&
\bigl[
u(\widetilde B_i,\widetilde c_i)
-u(\widetilde B_{i+1},\widetilde c_i)
\bigr]\\
&+
(1-\delta)
\bigl[
u(\widetilde B_{i+1},\widetilde c_i)
-u_{\min}
\bigr]\\
&+
\delta
\bigl(
\widetilde p_{i+1}-u_{\min}
\bigr)
\ge0.
\end{align*}
The first term is nonnegative because
$\widetilde B_i\supseteq\widetilde B_{i+1}$, the second is
nonnegative because every feasible dataset contains
$[\overline c,\overline v]$ and
$\widetilde c_i\le\overline c$, and the third is nonnegative by the
induction hypothesis.

We now define a continuous path in cutoff space. For
$s\in[i-1,i]$, where $i=1,\ldots,n-1$, set
\begin{equation}
\label{eq:cutoff-path}
\widetilde c_j(s)
=
\begin{cases}
\underline c,
& j<i,\\[3pt]
c_i-(s-i+1)(c_i-\underline c),
& j=i,\\[3pt]
c_j,
& j>i.
\end{cases}
\end{equation}
This path begins at the Step~1 cutoff vector and ends with every
cutoff equal to $\underline c$. Let
\[
V(s)
:=
\widetilde\Pi_1\bigl(\widetilde{\mathbf c}(s)\bigr)
\]
denote the broker's total payoff along the path. The recursive prices,
group masses, and continuation payoffs are continuous in the cutoff
vector, so $V$ is continuous. Moreover,
\[
V(0)=\Pi^{\mathrm{high}},
\qquad
V(n-1)=\delta^{n-1}u_{\min}<u_{\min}.
\]

We next verify the broker's incentive constraints at every point
$s$ for which
\begin{equation}
\label{eq:path-above-punishment}
V(s)\ge u_{\min}.
\end{equation}
Suppose that $s$ lies in phase $i$, and write
\[
x:=\widetilde c_i(s)\in[\underline c,c_i].
\]

At every stage $j\le i$, the active type set is the entire interval
$[\underline c,\overline c]$, because all earlier groups have zero
mass. No positive mass trades between stages $j$ and $i-1$, and hence
\[
\widetilde\Pi_j(s)
=
\delta^{\,i-j}\widetilde\Pi_i(s)
=
\delta^{\,1-j}V(s)
\ge
V(s)
\ge
u_{\min}.
\]
The punishment payoff after a deviation at any such stage is at most
$u_{\min}$, so the broker has no profitable deviation.

At every stage $j\ge i+2$, the active lower endpoint, the current
offer, all later offers, and all continuation group masses coincide
with those in the Step~1 equilibrium. Therefore,
\[
\widetilde\Pi_j(s)=\Pi_j^0,
\]
and the broker's incentive constraint holds by the choice
$\delta\ge\underline\delta_1(n)$ in Step~1.

It remains to verify stage $i+1$. This is precisely the stage omitted
by the earlier argument: its offer and all subsequent offers are
unchanged, but its active set begins at $x$ rather than at $c_i$.
Consequently,
\begin{equation}
\label{eq:modified-stage-i-plus-one}
\widetilde\Pi_{i+1}(s)
=
\Pi_{i+1}^0
+
\bigl[G(c_i)-G(x)\bigr]p_{i+1}^0.
\end{equation}
The punishment payoff at this history is
\[
u_{\min}\bigl[1-G(x)\bigr].
\]
Subtracting it from \eqref{eq:modified-stage-i-plus-one} gives
\begin{align*}
&\widetilde\Pi_{i+1}(s)
-u_{\min}\bigl[1-G(x)\bigr]\\
&\quad=
\left\{
\Pi_{i+1}^0
-u_{\min}\bigl[1-G(c_i)\bigr]
\right\}
+
\bigl[G(c_i)-G(x)\bigr]
\bigl(p_{i+1}^0-u_{\min}\bigr)
\ge0.
\end{align*}
The first term is nonnegative by the stage-$(i+1)$ incentive
constraint in Step~1. The second is nonnegative by
\eqref{eq:prices-above-umin}. Thus the broker's incentive constraint
also holds at stage $i+1$.

We have therefore proved that the strategy profile associated with
$\widetilde{\mathbf c}(s)$ is an equilibrium whenever
$V(s)\ge u_{\min}$. Define
\[
\overline s
:=
\inf
\left\{
s\in[0,n-1]:
V(s)\le u_{\min}
\right\}.
\]
Because
\[
V(0)=\Pi^{\mathrm{high}}>u_{\min}
>
V(n-1),
\]
the set is nonempty, $\overline s\in(0,n-1)$, and continuity implies
\[
V(\overline s)=u_{\min}.
\]
Moreover,
\[
V(s)\ge u_{\min}
\qquad
\text{for every }s\in[0,\overline s].
\]
Hence every point of this truncated path describes an equilibrium.

Finally, $V$ is continuous on $[0,\overline s]$, with
\[
V(0)=\Pi^{\mathrm{high}},
\qquad
V(\overline s)=u_{\min}.
\]
The intermediate value theorem therefore implies that every payoff in
\[
[u_{\min},\Pi^{\mathrm{high}}]
\]
is an equilibrium payoff. Since
$\Pi^{\mathrm{high}}\ge\pi(G)-\epsilon$, it follows that
\[
[u_{\min},\pi(G)-\epsilon]
\subseteq
SE(\delta).
\]
\end{enumerate}
\end{proof}

\subsection{Proofs of the Results in \cref{sec:discussion}}
\label{sec:additional}
The following lemma applies to the two-type quality model of \cref{prop:unique} with discount factor $\delta \in [0,1)$.
\begin{lemma}[Finite-Time Market Clearing]
\label{lem:ftmc}
There exists a finite $T(\delta)$ such that, in any equilibrium and after any history $h_t \in H_t$, the market clears within $T(\delta)$ periods along the equilibrium path.
\end{lemma}

\begin{proof}[Proof of \cref{lem:ftmc}]
Let $P(h_t)$ denote the remaining mass of buyers at history $h_t$. We show that for any equilibrium $(\sigma, \alpha)$ and any history $h_t \in H_t$ at which the market has not yet cleared, there exist a positive integer $\kappa(\delta)$ and a constant $k \in (0,1)$ such that
\[
P(h_{t+\kappa(\delta)}) \;\leq\; k \, P(h_t).
\]
Let $\overline{u}(\theta) = \theta x^e(\theta)-\frac{x^e(\theta)^2}{2}$. At history $h_t$, the seller can guarantee a payoff of at least
\[
P(h_t) \, u_{\min}
\]
by clearing the market immediately. Suppose that after $\kappa(\delta)$ periods a proportion $k$ of the mass $P(h_t)$ remains active. The seller's total expected profit is then bounded above by
\[
(1-k) P(h_t) \, \overline{u}(\theta_1)
\;+\; \delta^{\kappa(\delta)} k \, P(h_t) \, \overline{u}(\theta_1).
\]
Sequential rationality therefore requires
\[
\big[(1-k)+\delta^{\kappa(\delta)}k\big] \, \overline{u}(\theta_1)
\;\geq\; u_{\min} > 0.
\]
Thus, there exist $k$ and $\kappa(\delta)$ such that the remaining mass must contract by at least the factor $k$ within $\kappa(\delta)$ periods.

To complete the proof, note that the prior distribution contains a point mass at $\theta_2$. The geometric decay must eventually reach a state in which the remaining mass consists only of type $\theta_2$. At that point, the seller clears the market immediately.
\end{proof}

\begin{proof}[Proof of \cref{prop:unique}]
\cref{lem:ftmc} allows us to argue by backward induction.

The state can be represented by a variable $q \in (0,1]$, which corresponds to the remaining mass of buyers. Let $\theta(q)$ denote the left-continuous function mapping the state $q$ to the highest remaining buyer type. At the beginning of period $t$, we denote the state by $q^t$.

The structure of the proof follows \citet{gul1986foundations} and \citet{deneckere2006bargaining}.

Consider first the case in which $q^{t} \leq q_2$. Then $\theta(q^{t}) = \theta_2$, and the seller's optimal strategy is to clear the market immediately by offering
\[
x_t = x^e(\theta_2),
\qquad
p_t = u(x_t, \theta_2).
\]
Now consider a state $q^{t}$ satisfying $q_2 < q^{t} \leq 1$. In the final period $t = T(\delta)$, the seller clears the market by offering $(x_t, p_t)$ such that
\[
x_t = x^e(\theta_2),
\qquad p_t = u(x_t, \theta_2).
\]

Next, consider the penultimate period $t = T(\delta)-1$. Suppose the seller posts an offer $(x_t, p_t)$. We show that the resulting buyer behavior is uniquely determined. First, if
\[
u(x_t,\theta_2) \geq p_t,
\]
then the market clears immediately. Otherwise, type $\theta_2$ rejects, and type $\theta_1$ compares the current payoff with the continuation value. If
\[
u(x_t,\theta_1) - p_t > \delta [u(x^e(\theta_2),\theta_1) - u(x^e(\theta_2),\theta_2)],
\]
then all buyers of type $\theta_1$ purchase immediately; if the inequality is reversed, they wait.

A subtlety arises when this condition holds with equality. Although type $\theta_1$ is indifferent, equilibrium requires purchase with probability one. If a strictly positive mass of type $\theta_1$ were to wait, the seller could profitably deviate by offering a marginally lower price $p_t - \epsilon$, which induces immediate purchase by all of type $\theta_1$. Therefore, in equilibrium, indifference is resolved in favor of trade.

The seller's optimization problem is therefore well defined. Given the buyer's best response, the decision reduces to a comparison between screening type $\theta_1$ and clearing the market immediately.

For any $\delta \in [0,1)$, there exists a threshold mass $\bar{q}(\delta)$. If the remaining mass of buyers satisfies $q^t > \bar{q}(\delta)$, it is strictly profitable to screen type $\theta_1$ by offering $x_t = x^e(\theta_1)$. If $q^t < \bar{q}(\delta)$, the cost of delaying sales to the larger mass of type $\theta_2$ dominates the screening benefit, and the seller strictly prefers to clear the market immediately. When $q^t = \bar{q}(\delta)$, the seller is indifferent.

Indifference at $\bar{q}(\delta)$ is what guarantees existence. On path, type $\theta_1$ can randomize so that the posterior mass equals exactly $\bar{q}(\delta)$. Off path, after a seller deviation, the seller can randomize between screening and clearing at states where he is indifferent, which convexifies the buyers' continuation payoffs and allows play to return to the equilibrium path. Together, these ensure that an equilibrium exists after every history.

Proceeding by backward induction on the horizon, we characterize the equilibrium path. It consists of a sequence of periods in which the seller offers $x^e(\theta_1)$ and screens type $\theta_1$ through delay, followed by a final period that clears type $\theta_2$. By monotone comparative statics \citep{MilgromShannon}, the equilibrium path is unique and deterministic, with the possible exception of randomization in the initial period; this is the sense in which the equilibrium outcome is essentially unique in \cref{prop:unique}. After a seller deviation, existence is guaranteed by the seller's randomization described above, which returns play to the equilibrium path. Moreover, the equilibrium is weak-Markov. As these results are standard in the literature, we omit the detailed derivation.

Finally, we establish a bound $T$ on the length of the equilibrium path that is independent of $\delta$. This follows from the presence of a point mass at $\theta_2$. Accelerating trade, which is feasible under weak-Markov strategies, yields a benefit bounded below by $(1-\delta)L$ for some constant $L > 0$. On the other hand, the price increment used for screening is constrained by
\[
p_t - p_{t+1} \leq p_t - \delta p_{t+1} = (1-\delta) u(x^e(\theta_1),\theta_1).
\]
After normalizing by the factor $(1-\delta)$, the optimality condition implies that a positive mass of buyers must clear in each period. This requirement yields a bound $T$ that is independent of $\delta$. Consequently, as $\delta \to 1$, the total discounting over the horizon, $\delta^T$, converges to $1$. The resulting outcome combines efficient qualities with vanishing delay.
\end{proof}

\end{document}